\documentstyle[12pt]{article}
\newcommand{\newsection}{
\setcounter{equation}{0}
\section}

\def\appendix#1{
  \addtocounter{section}{1}
  \setcounter{equation}{0}
  \renewcommand{\thesection}{\Alph{section}}
 \section*{Appendix \thesection\protect\indent \parbox[t]{11.715cm} {#1}}
  \addcontentsline{toc}{section}{Appendix \thesection\ \ \ #1}
  }

\newcommand{\tr}[1]{\:{\rm tr}\,#1}
\newcommand{\Tr}[1]{\:{\rm Tr}\,#1}
\def\e{\,{\rm e}\,}
\newcommand{\rf}[1]{(\ref{#1})}

\newcommand{\non}{\nonumber \\*}
\def\be{\begin{equation}}
\def\ee{\end{equation}}
\def\bea{\begin{eqnarray}}
\def\eea{\end{eqnarray}}
\def\const{{\rm const}}
\def\ll{\Lambda}
\def\bp{\bar{\varphi}}
\def\ba{\bar{A}}
\def\bd{\bar{D}}
\def\bs{\bar{F}}
\def\bm{\bar{\Pi}}
\def\d{\partial}
\def\D{\delta}
\def\a{\alpha}
\def\b1{{\bf 1}}

\begin{document}

\begin{titlepage}
\begin{flushright}
hep-th/9803237\\
March, 1998
\end{flushright}
\vspace{1.5cm}

\begin{center}
{\LARGE Renormalization of Functional} 
\\[.5cm]
{\LARGE Schr\"odinger Equation}
\\[.5cm]
{\LARGE by Background Field Method}\\
\vspace{1.9cm}
{\large K.~Zarembo}\\
\vspace{24pt}
{\it Department of Physics and Astronomy,}
\\{\it University of British Columbia,}
\\ {\it 6224 Agricultural Road, Vancouver, B.C. Canada V6T 1Z1} 
\\ \vskip .2 cm
and\\ \vskip .2cm
{\it Institute of Theoretical and Experimental Physics,}
\\ {\it B. Cheremushkinskaya 25, 117259 Moscow, Russia} \\ \vskip .5 cm
E-mail: {\tt zarembo@theory.physics.ubc.ca/@itep.ru}
\end{center}
\vskip 2 cm
\begin{abstract}
Renormalization group transformations for Schr\"odinger 
equation are performed in $\varphi^4$ and in Yang-Mills 
theories. The dependence of the ground state wave functional 
on rapidly oscillating fields is found. For Yang-Mills theory, 
this dependence restricts a possible form of variational 
ansatze compatible with asymptotic freedom.  
\end{abstract}

\end{titlepage}
\setcounter{page}{2}

\newsection{Introduction}

It is widely accepted that many physical properties of  
gauge theories in the confining phase are determined by the structure 
of their vacua. In the literal sense of the term, the vacuum is 
described by a ground state wave functional -- the lowest energy 
solution of the Schr\"odinger equation. To find a complete solution
of the Schr\"odinger equation in Yang-Mills theory is by no means 
feasible, but the asymptotic freedom should allow to determine
the dependence of the ground state wave functional on rapidly 
oscillating fields, since this dependence is responsible for the 
renormalization.

General properties of the renormalization of functional 
Schr\"odinger equation were studied  within the perturbation theory.
It was shown that both the Hamiltonian and the wave functional
are renormalizable by usual counterterms up to multiplicative 
redefinition of the field variables \cite{sym81}. Practically, the 
most convenient way to perform renormalization group transformations 
explicitly is based on the averaging over rapidly oscillating
degrees of freedom. Background field method in the path integral
framework \cite{dew65} is usually associated with this procedure.
The Hamiltonian counterpart of the background field method 
was also employed in some problems, for example,
in connection with gauge fields on a torus \cite{lus83,smi87}, 
in soliton quantization \cite{raj89} and also in variational
calculations in gauge theories \cite{hmvi98}.
   
The averaging procedure is common in quantum mechanics with
finite number of degrees of freedom. An instructive example
\cite{sim83,med84} is the system with  
Hamiltonian 
\be\label{hsim}
H=\frac{1}{2}\,p_x^2+\frac{1}{2}\,p_y^2+\frac{1}{2}\,x^2y^2.
\ee
The potential energy for this system 
is degenerate along the coordinate
axes. Far from the origin, say, at $|x|\gg 1$, the potential
valley becomes very narrow and the wave function varies along 
the valley much slower than in the transverse direction.
So, the variables are naturally separated in the slow and the fast
ones. The last two terms in \rf{hsim} is the
Hamiltonian for the fast mode $y$: 
$H_h=\frac{1}{2}\,p_y^2+\frac{1}{2}\,x^2y^2$. 
Its ground state energy induces 
an effective potential for the slow degree of freedom: 
$V_{\rm eff}(x)=|x|/2$. Thus, the degeneracy of the potential is 
lifted and the system
with Hamiltonian \rf{hsim}, in fact, has a discrete spectrum 
\cite{sim83}. The above procedure can be generalized to a field 
theory \cite{lus83}, where the problem of lifting of the classical
vacuum degeneracy also can be addressed \cite{lus83,kmo97}.
In this paper we use the averaging procedure in a 
systematic way to renormalize Schr\"odinger equation in 
$\varphi^4$ and in the Yang-Mills theories.
We consider pure gauge theory mostly for the sake of 
simplicity and the introduction of quarks should not cause 
any difficulties, since suitable methods to treat fermions in the
functional Schr\"odinger picture are known \cite{fj88}.

It is worth mentioning that the averaging over fast modes in not
the only way to renormalize Schr\"odinger equation.
As usual, the renormalization group transformations consist in   
a proper modification of the 
Hamiltonian for low-energy degrees of freedom after elimination of
the fast modes \cite{wil65}. So,
the renormalized Hamiltonian acts in the smaller Hilbert space
than the bare one and the renormalization  can be
considered as a projection on this smaller space, 
or a reduction of the Hamiltonian to the 
block-diagonal form. General methods for partial diagonalization
\cite{gw94,weg94}, as well as for projection \cite{fgv95} of 
Hamiltonians exist. These methods were tested on simple 
quantum-mechanical systems \cite{vp92,fgv95} and applied to some
many-body \cite{weg94,bm3} and field-theoretical light-cone \cite{bm1} 
and equal-time \cite{bm2} Hamiltonians.

\newsection{Scalar field}

The Hamiltonian of $\varphi^4$ theory is
\be\label{phi4}
H=\int d^3x\,\left[\frac{1}{2}\,\Pi^2+\frac{1}{2}\,
(\partial\varphi)^2
+\frac{1}{2}\,m^2\varphi^2+\frac{1}{24}\,\lambda
\varphi^4\right],
\ee
where $\Pi$ is canonically conjugate to the field variable:
\be
[\varphi(x),\Pi(y)]=i\delta(x-y).
\ee
Schr\"odinger equation for the Hamiltonian \rf{phi4} requires
regularization, thus we assume that modes with momenta larger than
$\ll$ are somehow excluded (the concrete prescription will be given
below). In order to exclude also the modes with momenta larger than 
$\mu$, where $\ll\gg\mu\gg m$, we can explicitly solve 
Schr\"odinger equation
for the fields containing only these modes and then average
the Hamiltonian with the wave function for the high-energy degrees
of freedom.
Denote by $\bp$, $\bm$ and by $\phi$, $\pi$ the slow and the fast 
variables -- the field components which contain  
modes with momenta $p<\mu$ and $\mu<p<\ll$, respectively. 
According to the conventional assumption
of the background field formalism \cite{dew65}, slowly varying 
fields $\bp$ satisfy classical equations of motion.

Extracting the Hamiltonian for the high-energy degrees of freedom
and expanding it to the quadratic order in $\phi$ (which is 
equivalent to the lowest order perturbation theory in $\lambda$), we
get:
\be\label{phi4h}
H_h=\int d^3x\,\left[\frac{1}{2}\,\pi^2+\frac{1}{2}\,
(\partial\phi)^2
+\frac{1}{2}\,m^2\phi^2+\frac{1}{4}\,\lambda
\bp^2\phi^2\right].
\ee
Since this Hamiltonian is quadratic, the ground state wave function
has the form
\be
\Psi_h=\exp\left(-\frac{1}{2}\,
\int d^3xd^3y\,\phi(x)K(x,y)\phi(y)\right).
\ee
Substituting this expression in the Schr\"odinger equation
\be
\int d^3x\,\left[-\frac{1}{2}\,\frac{\D^2}{\D\phi^2}+\frac{1}{2}\,
(\partial\phi)^2
+\frac{1}{2}\,m^2\phi^2+\frac{1}{4}\,\lambda
\bp^2\phi^2\right]\Psi_h=E_h\Psi_h,
\ee
 we obtain for the operator $K$:
\be
K=\left(-\d^2+m^2+\frac{1}{2}\,\lambda\bp^2\right)^{1/2},
\ee
the ground state energy being equal to
\be\label{grh}
E_h=\frac{1}{2}\,\Tr K.
\ee

As in the quantum-mechanical example in the introduction,
the ground state energy of large-momentum modes renormalizes the 
low-energy Hamiltonian. When $\mu$ is much larger than any other
scale in the problem, only the ultraviolet divergent contributions
are  essential. UV divergences can be easily
extracted by the heat kernel method, which is based on the following
representation for the square root of the operator: 
\be\label{root}
K=\lim_{\varepsilon\rightarrow 0}\left(\frac{1}{2\varepsilon}
-\frac{1}{2\pi^{1/2}}\,\int_{0}^{\infty}\frac{d\tau}{\tau^{3/2}}
\,\e^{-\tau K^2-\frac{\varepsilon^2}{\tau}}\right).
\ee
When we calculate the trace, the field-independent term  
merely renormalize the zero-point energy,
$$E_0=\int\frac{d^3p}{(2\pi)^3}\,\frac{\sqrt{p^2+m^2}}{2}.$$
Below this field-independent contribution is ignored. The 
remaining part of the effective potential can be calculated,
for example, expanding 
\be\label{vhk}
V_{\rm eff}(\bp)\equiv E_h=\const
-\frac{1}{4\pi^{1/2}}\,\int\frac{d\tau}{\tau^{3/2}}\,
\Tr\exp\left[-\tau\left(-\d^2+m^2+\frac{1}{2}\,\lambda\bp^2\right)
\right]
\ee
in $\lambda$. This procedure involves evaluation of the momentum
integrals, where the momentum cutoff can be introduced as a lower
bound of integration over $\tau$. Therefore, the prescription to
integrate over $\tau$ from $\ll^{-2}$ to $\mu^{-2}$ naturally
accounts for the 
presence of only large-momentum modes in $\phi$.

In fact, direct expansion in $\lambda$ is not the most convenient
way to extract the essential terms in the effective potential.
The UV divergences are governed by the behavior of the integrand
in \rf{vhk} at small $\tau$ and can be easily captured by  
DeWitt-Seeley expansion \cite{dew65}. In particular \cite{sch89}:
\be\label{phi4ds}
\langle x |\e^{-\tau(-\d^2+V)}| x\rangle
=\frac{1}{(4\pi\tau)^{3/2}}\left[1+V\tau+\left(\frac{1}{2}\,
V^2-\frac{1}{6}\,\d^2V\right)\tau^2+O(\tau^3)\right].
\ee
Substituting $\lambda\bp^2/2$ for $V$, we get from 
eqs.~\rf{vhk}, \rf{phi4ds}:
\be\label{phi4veff}
V_{\rm eff}(\bp)=\const-\frac{1}{4\pi^{1/2}}\,\int
\frac{d\tau}{\tau^{3/2}}\,\frac{\e^{-m^2\tau}}{(4\pi\tau)^{3/2}}
\int d^3x\,\left(\frac{1}{2}\,\lambda\bp^2\tau
+\frac{1}{8}\lambda^2\bp^4\tau^2\right)+O(1/\mu^2),  
\ee
The divergent terms lead to  quadratic renormalization of the
mass and logarithmic renormalization of the coupling:
\be
\lambda_{\rm eff}=\lambda-\frac{3\lambda^2}{32\pi^2}\,
\ln \frac{\ll^2}{\mu^2}.
\ee
This result, of course, gives the correct expression for the 
$\beta$-function in the $\varphi^{4}$ theory.

\newsection{Yang-Mills theory}

The Hamiltonian formulation of the Yang-Mills theory is most simple
in the temporal gauge $A_0=0$. Then the canonical variables are
gauge potentials $A_i^B(x)$ and electric fields $E_i^B(x)$
(we consider $SU(N)$ gauge group, so $B=1,\ldots,N^2-1$):
\begin{equation}\label{ccr}
[A_i^A(x),E_j^B(y)]=i\delta ^{AB}\delta _{ij}\delta (x-y).
\end{equation}
The Hamiltonian is
\begin{equation}\label{ham}
H=\int d^3x\,\left(\frac{g^2}{2}\,E_i^AE_i^A
+\frac{1}{4g^2}\,F_{ij}^AF_{ij}^A\right),
\end{equation}
where $F_{ij}^A=\partial _iA_j^A-\partial _jA_i^A+f^{ABC}A_i^BA_j^C$.
The wave functions of physical states are also subject to the Gauss'
law constraint:
\begin{equation}\label{gauss}
D_iE_i^A\, \Psi =0.
\end{equation}
The covariant derivative $D_i$ acts in the adjoint representation:
$D_i^{AB}=\delta ^{AB}\partial _i+f^{ACB}A_i^C$.

In the ``coordinate'' representation, 
the wave function is a functional of the gauge potentials and
the electric field operators act as
the variational derivatives: $E_i^A(x)=-i\,\delta /\delta A_i^A(x)$.
In order to find the dependence of the vacuum wave functional on 
the field modes with high momenta, we proceed in the same way as
in the previous section. The slowly varying fields, 
 $\ba_i$, are assumed to contain only modes with momentum
 $p<\mu$ and to satisfy classical equations of motion:
\be\label{cl}
\bd_i\bs_{ij}=0,
\ee
where $\bd_i$ is the covariant derivative with respect to $\ba_i$
and $\bs_{ij}$ is the corresponding field strength.
It is convenient to rescale the fast variables $a_i$, which contain
the modes with momenta $\mu<p<\ll$, so that $A_i=\ba_i+ga_i$.
To preserve the canonical commutation relations for the fast modes, 
the high-energy components of the electric fields should 
be rescaled by $1/g$: $E_i=\bar{E}_i+\frac{1}{g}\,e_i$.

We assume that the scale $\mu$ is sufficiently large, so that the 
running coupling $g(\mu)$ is small, and we can use perturbation
theory for high-energy Hamiltonian. Expanding
the Hamiltonian for rapidly oscillating variables in $g$ to the
leading order, we get:
\be\label{ymham}
H_h=\int d^3x\,\left[\frac{1}{2}\,e_i^Ae_i^A+\frac{1}{2}\,
a_i^A(-\bd^2\D_{ij}-2\bs_{ij}+\bd_i\bd_j)^{AB}a_j^B\right].
\ee
Here $\bs_{ij}^{AB}=[\bd_i,\bd_j]^{AB}=f^{ACB}\bs_{ij}^C$ acts as
a matrix in the adjoint representation.

Denote by $L$ the quadratic form of the potential in eq.~\rf{ymham}: 
\be\label{qf}
L_{ij}=-\bd^2\D_{ij}-2\bs_{ij}+\bd_i\bd_j,
\ee
Then the ground state wave functional for the fast variables is
\be\label{ymwf}
\Psi_h=\exp\left(-\frac{1}{2}\,aL^{1/2}a\right),
\ee
where summation over color and spatial indices and integration over
spatial coordinates is implied. The wave functional \rf{ymwf} 
satisfies Gauss' law constraint up to the two first orders in $g$:
\be
\left(\frac{1}{g}\,\bd_ie_i^A+\bd_i\bar{E}_i^A+f^{ABC}\ba_i^Be_i^C
\right)\Psi_h=0.
\ee
The $O(1/g)$ part of the Gauss' law generates transformations
\be\label{g1}
a_i\rightarrow a_i+\bd_i\omega.
\ee
The invariance of the wave functional under these transformations 
follows 
from the identity $L_{ij}\bd_j\omega=0$ \cite{hmvi98}. Terms of
order $g^0$ in the Gauss' law generate usual 
gauge transformations of the
background fields and homogeneous transformations of the fast 
variables:
\be\label{g2}
\ba_i\rightarrow\Omega^{\dagger}(\d_i+\ba_i)\Omega,
~~~~a_i\rightarrow\Omega^\dagger a_i\Omega.
\ee
The invariance of $\rf{ymwf}$ under these transformations is evident,
since the kernel of the operator $L^{1/2}$ transforms homogeneously:
$L^{1/2}(x,y)\rightarrow\Omega^\dagger(x) L^{1/2}(x,y)\Omega(y)$.
To recover the invariance of the wave functional
under the transformations generated by higher order terms in the 
Gauss' law, it is necessary to include higher orders in the 
Schr\"odinger equation.

The effective potential generated by averaging 
of the Hamiltonian with the wave functional
\rf{ymwf} is 
\be\label{ymveff}
V_{\rm eff}(\ba)=\frac{1}{2}\,\Tr L^{1/2}.
\ee
Although the operator $L$ has a large number of gauge zero modes,
they do not cause any problems with the calculation of the trace of its
square root, hence, there is no reason to fix the gauge and to introduce
ghosts as in the conventional background field method, where one usually
deals with logarithms of the operators.
The field-independent part of the effective potential
corresponds to the zero-point energy and 
is not traced below. If the scale $\mu$ is sufficiently large, only
UV divergent contributions are essential. The only contribution of
this kind, the gauge coupling renormalization, can be easily extracted 
by the heat kernel method, as in Ref.~\cite{hmvi98}, where a similar 
expression was considered 
in the context of the variational approach to the QCD vacuum.

DeWitt-Seeley coefficients for
the operator $L$ were calculated in Ref.~\cite{glno93}, where a more 
general operator,
\be\label{qfa}
L_{ij}(\a)=-\bd^2\D_{ij}-2\bs_{ij}+\left(1-\frac{1}{\a}
\right)\bd_i\bd_j,
\ee
was considered. Taking the limit $\a\rightarrow\infty$ of the 
small $\tau$ expansion
of the heat kernel given in Ref.~\cite{glno93}\footnote{In fact, 
the first, field-independent, 
DeWitt-Seeley coefficient of the operator $L(\protect\a)$ diverges as 
$\protect\a\rightarrow\protect\infty$. Although we do not trace the 
field-independent
contribution, the origin of this divergence is clarified in Appendix.
The correct prescription consists in simply dropping the divergent 
term.}, we find:
\bea
V_{\rm eff}(\ba)&=&\const-\frac{1}{4\pi^{1/2}}\,
\int\frac{d\tau}{\tau^{3/2}}\,
\int d^3x\,\tr\langle x|\e^{-\tau L}|x\rangle
\non
&=&-\frac{1}{4\pi^{1/2}}\,\int\frac{d\tau}{\tau^{3/2}}\,
\frac{1}{(4\pi\tau)^{3/2}}\,\left(\const+
\tau^2\,\frac{11N}{6}\int d^3x\,
\bs_{ij}^A\bs_{ij}^A+O(\tau^3)\right)
\non
&=&\const-\frac{11N}{48\pi^2}\,\ln\frac{\ll^2}{\mu^2}\,\frac{1}{4}\,
\int d^3x\,\bs_{ij}^A\bs_{ij}^A+O(1/\mu^2).
\eea
Here $\tr$ denotes the trace with respect to 
color and to spatial indices.
The averaging over the fast modes reproduces the usual coupling
constant renormalization:
\be
\frac{1}{g^2_{\rm eff}}=\frac{1}{g^2}-\frac{11N}{24\pi^2}\,
\ln\frac{\ll}{\mu}\,,
\ee
as expected.

\newsection{Discussion}

The most common application of the background field 
techniques, that to QCD sum rules \cite{nsvz84}, is based on the 
parametrization of the slowly varying
fields by vacuum condensates \cite{svz79}. In the 
Hamiltonian picture,
this would correspond to treating rapidly oscillating fields 
perturbatively, as above, and parameterizing the wave functional for 
the low-energy degrees of freedom by a finite number of parameters.    
Probably, this approach may provide some useful information.

One may try to parametrize the vacuum in Yang-Mills theory
by a finite number of parameters using some reasonable ansatz for 
the ground state wave function and then to find
an approximate solution of the Schr\"odinger equation from
the variational principle (for a recent discussion 
of the variational approach to the QCD vacuum, see 
\cite{kk95,hmvi98}). Apparently, an accuracy of such kind of
approximations is not under control. In particular, the correct 
renormalization of physical quantities is not
automatically guaranteed in the variational calculations.
Our results show that any approximate 
wave functional compatible with 
the asymptotic freedom must depend on the large-momentum modes
of the fields in a very specific way -- as given by eq.~\rf{ymwf}. 
This property substantially restricts valid functional form 
of variational 
ansatze  for the ground state in Yang-Mills theory.

\subsection*{Acknowledgments}

This work was supported by NATO Science Fellowship and, in part, by
 CRDF grant 96-RP1-253,
 INTAS grant 96-0524,
 RFFI grant 97-02-17927
 and grant 96-15-96455 of the support of scientific schools.

\setcounter{section}{0}
\setcounter{subsection}{0}
\appendix{Heat kernel of the operator $L$}

In order to relate the operators $L$ and $L(\a)$, it is convenient
to introduce the non-Abelian transverse projector
\cite{orl96,hmvi98}:
\be
P_{ij}=\D_{ij}-\bd_i\,\frac{1}{\bd^2}\,\bd_j.
\ee
Denote $L(1)$ by $G$: 
\be\label{qfg}
G_{ij}=-\bd^2\D_{ij}-2\bs_{ij}.
\ee
Then, with the use of the equations of motion \rf{cl} 
and the commutation 
relations $[\bd_i,\bd_j]=\bs_{ij}$, the following equality can 
be proved:  
\be\label{proj}
L_{ij}=G_{ik}P_{kj}=P_{ik}G_{kj},
\ee
So, the operator $L$ is a transverse projection of the operator $G$.

As a consequence of eq.~\rf{proj}, and since $P$ is the projection 
operator,
\be
\e^{-\tau L}=\e^{-\tau G}P.
\ee
One the other hand, eqs.~\rf{qf}, \rf{qfg} and \rf{proj} imply that
$\bd_i\bd_j=-G_{ik}(\D_{kj}-P_{kj})$ and, consequently,
\be
L_{ij}(\a)=G_{ik}P_{kj}+\frac{1}{\a}\,G_{ik}(\D_{kj}-P_{kj}).
\ee
Since $P$ and $\b1-P$ are the orthogonal projectors,
\be\label{hka}
\e^{-\tau L(\a)}=\e^{-\tau G}P+\e^{-\frac{\tau}{\a}\,G}(\b1-P).
\ee

The diagonal matrix elements of the second term on the right hand
side of eq.~\rf{hka} can be expanded in local operators. By 
dimensional reasons the parameter of this expansion is $\tau/\a$:
\be
\tr\langle x |\e^{-\frac{\tau}{\a}\,G}(\b1-P) | x \rangle
=\sum_{n=0}^{\infty}\left(\frac{\tau}{\a}\right)^{n-\frac{3}{2}}
{\cal O}_n(x),
\ee
where $\tr$ denotes the trace with respect to spatial and to color 
indices and ${\cal O}_n(x)$ are gauge-invariant operators of
dimension $2n$. As there are no such operators of dimension $2$,
 only the first term proportional 
to $\a^{3/2}$ survives the limit $\a\rightarrow\infty$. 
It is field-independent and can be 
calculated in the momentum representation. Finally, we get:
\be
\tr\langle x|\e^{-\tau L}|x\rangle=\lim_{\a\rightarrow\infty}\left[
\tr\langle x|\e^{-\tau L(\a)}|x\rangle -
\frac{(N^2-1)\a^{3/2}}{(4\pi\tau)^{3/2}}\right].
\ee
The small $\tau$ expansion of the right hand side is given in 
Ref.~\cite{glno93} up to the two first terms and, indeed, has a
finite $\a\rightarrow\infty$ limit.

\end{document}